\documentclass[usenatbib]{mn2e}
\usepackage{graphicx,times}
\usepackage{bm,url}
\newcommand{\EQ}{\begin{equation}}
\newcommand{\EN}{\end{equation}}
\newcommand{\EQA}{\begin{eqnarray}}
\newcommand{\ENA}{\end{eqnarray}}
\newcommand{\nab}{\nabla}

\newcommand{\meanB}{\overline{B}}

\newcommand{\meanemf}{\overline{\cal E} {}}
\newcommand{\meanEMF}{\overline{\mbox{\boldmath ${\mathcal E}$}} {}}

\newcommand{\meanBB}{\bm{\overline{{B}}}}

\newcommand{\meanUU}{\bm{\overline{{U}}}}
\newcommand{\oo}{\bm{{{\omega}}}}
\newcommand{\uu}{\bm{{{u}}}}
\newcommand{\bb}{\bm{{{b}}}}
\newcommand{\ff}{\bm{{{f}}}}
\newcommand{\kk}{\bm{{{k}}}}

\newcommand{\GG}{\bm{{{G}}}}

\def\const{{\rm const}}

\def\Pm{P_\mathrm{m}}
\def\Rm{R_\mathrm{m}}
\def\Rey{\mbox{\rm Re}}

\def\onethird{{\textstyle{1\over3}}}

\newcommand{\MMMM}{\mbox{\boldmath ${\sf M}$} {}}

\newcommand{\SSSS}{\mbox{\boldmath ${\sf S}$} {}}

\newcommand{\Eq}[1]{equation~(\ref{#1})}

\newcommand{\Fig}[1]{Fig.~\ref{#1}}

\title[Kinematic alpha effect]%
{Kinematic alpha effect in isotropic turbulence simulations}
\author[S.~Sur, A.~Brandenburg, K.~Subramanian]%
{Sharanya Sur$^{1}$, Axel Brandenburg$^2$ and Kandaswamy Subramanian$^{1}$
\thanks{E-mail: sur@iucaa.ernet.in (SS); brandenb@nordita.dk (AB);
kandu@iucaa.ernet.in (KS)}\\
$^{1}$Inter-University Centre for Astronomy and
        Astrophysics,  Post Bag 4, Ganeshkhind, Pune 411 007, India\\
$^2$NORDITA, AlbaNova University Center, SE - 106 91 Stockholm, Sweden}

\date{}
\begin{document}

\pagerange{\pageref{firstpage}--\pageref{lastpage}} \pubyear{2006}

\maketitle

\begin{abstract}
Using numerical simulations
at moderate magnetic Reynolds numbers up to 220
it is shown that in the kinematic regime,
isotropic helical turbulence leads to an alpha effect and a turbulent
diffusivity whose values are independent of the magnetic Reynolds number,
$\Rm$, provided $\Rm$ exceeds unity.
These turbulent coefficients are also consistent with expectations
from the first order smoothing approximation.
For small values of $\Rm$, alpha and turbulent diffusivity are
proportional to $\Rm$.
Over finite time intervals meaningful values of alpha and
turbulent diffusivity can be obtained even when there is small-scale
dynamo action that produces strong magnetic fluctuations.
This suggests that small-scale dynamo-generated fields 
do not make a correlated contribution to the mean electromotive force.
\end{abstract}
\label{firstpage}
\begin{keywords}
magnetic fields --- MHD --- hydrodynamics -- turbulence
\end{keywords}

\section{Introduction}
The generation and maintenance of large-scale magnetic fields
in stars and galaxies is often studied within
the framework of the mean-field dynamo (MFD);
see, e.g., \cite{Mof78,Par79,KR80}.
A particularly important driver of MFDs is the $\alpha$-effect.
For isotropic turbulence and weak magnetic fields, i.e.\ in
the kinematic regime, the $\alpha$-effect
can be expressed purely in terms of the kinetic helicity.
Research in recent years has mostly been concerned with clarifying
the effects of nonlinearity, but there are serious uncertainties
even in the linear (kinematic) regime.
In particular, whether or not $\alpha$ can then be expressed
in terms of the kinetic helicity depends on the
applicability of the first order smoothing approximation (FOSA)
or other closures used to calculate $\alpha$.
Such approaches become questionable
when the magnetic Reynolds number, $\Rm$, is large, i.e.\
when the magnetic diffusion time is long compared
with the turnover time which, in turn, is comparable with
the correlation time of the turbulence.
In the high conductivity limit FOSA can only be applied
if the correlation time for the
velocity field is much smaller than the eddy turnover time.
This is not the case for high Reynolds number turbulence, where
the two time scales are equal, i.e.\ the Strouhal number is unity
\citep{BS05b,BS07},
so FOSA should in principle break down.
In this case all higher order terms 
need to be taken into account \citep{Kno78}.
Furthermore, high $\Rm$ random flows typically lead to a fluctuation dynamo
which leads to rapidly growing small-scale magnetic fields independent of the
mean field. This also breaks the assumption made by 
FOSA that fluctuating fields are much smaller than the mean field.

The existence of $\alpha$-effect and turbulent diffusion has been
worrying dynamo researchers over several decades.
Assuming steady flow patterns, \cite{Chi79} found that motions that 
concentrate magnetic fields into thin flux sheets lead to
an $\alpha$-effect whose value diminishes with $\Rm$ like $\Rm^{-1/2}$.
But for an analogous helical motion which concentrates the field
into an axial flux rope, $\alpha$ tends to a finite limit as $\Rm \to \infty$.
He conjectured that the latter estimate may be typical of steady three-dimensional
motion.  The validity of turbulent diffusion has been questioned by \cite{Pid81}.
Calculations by \cite{Kraichnan76} suggest that
$\alpha$ and turbulent diffusion converge to finite values
for statistically isotropic velocity fields with gaussian statistics.
However, numerical simulations \citep{Drummond86} using a frozen velocity
field suggest that in the limit
of large magnetic Reynolds numbers $\alpha$ tends to zero.
Based on specific imposed (kinematic) flow patterns it has been
suggested that there is no simple relation between $\alpha$ and helicity
of the flow; see \cite{CHT06}.
In fact, their results may suggest that in the kinematic regime,
$\alpha$ exhibits a strong $\Rm$ dependence
(for large $\Rm$ up to $2\times10^5$)
and could change sign
for $\Rm\approx20$ and long correlation times.
It is important to emphasize that in the nonlinear regime, i.e.\ for finite
magnetic field strength and including the Lorentz force,
a strong $\Rm$ dependence is now indeed well established
\citep{CH96,B01}; see \cite{BS05a} for a review.
However, in the following we shall be concerned with the purely linear regime.

In order to clarify the $\Rm$ dependence in the kinematic regime
we perform numerical turbulence experiments where we 
adopt an externally imposed body force to drive the flow.
This is a common technique applied in simulations and helps develop
homogeneous and isotropic turbulence which is easier to handle analytically.
While the calculation of $\alpha$ from a turbulence simulation is
relatively straightforward by imposing a uniform magnetic field, the
calculation of turbulent diffusion is more uncertain.
One possibility is to determine the decay rate of an initial large-scale
magnetic field \citep{YBR03}.
Another more reliable method is to calculate both $\alpha$ and turbulent
diffusion tensors simultaneously by computing the mean electromotive force using
a number of different nonuniform test fields in different
directions and of different spatial structure \citep{S05,S07}.
This is also the approach used in the present paper.

\section{Test-field procedure}
In MFD theory,
one averages the induction equation to obtain
the standard dynamo equation for the
mean field $\meanBB$,
\EQ
\label{indmeanB}
{\partial{\meanBB}\over\partial t}=
\nab\times(\meanUU\times\meanBB+\meanEMF - \eta\nab \times \meanBB), \quad
\nab\cdot\meanBB=0.
\EN
This averaged equation now has a new term,
the mean electromotive force (emf)
$\meanEMF={\overline {\uu\times\bb}}$, 
which crucially depends on
the statistical properties of the small-scale velocity and magnetic
fields, $\uu$ and $\bb$, respectively.
A central closure problem in MFD theories is 
to compute the mean emf $\meanEMF$
and express it in terms of the mean field itself.
Assuming that the mean field is spatially
smooth, the mean emf $\meanEMF$ can then be expressed in terms of the 
mean magnetic field and its first derivative in a manner 
\EQ
\label{meanemf1}
\meanemf_i=\alpha_{ij}\meanB_{j} + \eta_{ijk}\meanB_{j,k},
\EN
where $\alpha_{ij}$ and $\eta_{ijk}$ are turbulent transport coefficients
written in tensorial form, and a comma denotes partial differentiation.

In numerical simulations, the full $\alpha_{ij}$ and $\eta_{ijk}$ tensors
are determined by 
first computing the mean emf using test fields in different directions
and of different spatial structure \citep{S05,S07}.
In what follows, we employ $\it{xy}$ averages such that the resulting
mean fields are expressible only as functions of $z$ and $t$.
Some details of the test field method applied to this case have already
been described by \cite{B05}.
Note that the solenoidality condition then gives, $\meanB_{z}=\const=0$.
Hence, one only needs to compute the four components of
$\alpha_{ij}$ and $ \eta_{ij3}$ with $i,j=1,2$.
Here the numbers 1, 2, and 3 refer to Cartesian coordinate directions
$x$, $y$, $z$.
In order to obtain the $4+4=8$ unknown coefficients, we need
the $x$ and $y$ components of 4 different test fields, $\meanB_i^{pq}$,
where $i$, $p$, and $q$ take values 1 and 2.

In order to compute the components $\alpha_{11}$ and $\eta_{123}$, for example,
it suffices to consider the following two test fields,
\EQ
\meanBB^{11}=\pmatrix{\cos k_1 z\cr0\cr0},\quad
\meanBB^{21}=\pmatrix{\sin k_1 z\cr0\cr0}.
\EN
Here $k_1$ is the smallest non-vanishing wavenumber in the domain,
and $p=1$ or $p=2$ denotes whether we take a cosine or sine behaviour
for the test field, and $q=1$ or $q=2$ depending on whether the non-zero
component of $\meanB_i^{pq}$ is the $x$-component or $y$-component respectively.
We insert these test fields into the relation
\EQ
\label{meanemf2}
\meanEMF_{i}^{pq}
={\alpha_{ij}}\meanB_{j}^{pq}+\eta_{ij3}\meanB_{j,3}^{pq}.
\EN
Since $q=1$, only the $j=1$ ($x$-) component contributes to the sum over $j$ above.
For the two values of $p$ and with index $i$ unspecified, we have
\EQA
\meanemf_i^{11}=\alpha_{i1}\cos k_1 z-\eta_{i13}k_1\sin k_1 z,\\
\meanemf_i^{21}=\alpha_{i1}\sin k_1 z+\eta_{i13}k_1\cos k_1 z.
\ENA
Similarly, for test fields which only have non-zero $y$-components, that is 
with $q=2$,
one obtains a similar pair of equations
with the same arrangement of cosine and sine functions,
but with $\alpha_{i2}$, $\eta_{i23}$ in place of $\alpha_{i1}$ and $\eta_{i13}$.
So, for each value of $i$ one obtains independent matrix equations
for the unknown coefficients $\alpha_{ij}$ and $\eta_{ij3}$ as
\EQ
\pmatrix{\alpha_{ij}\cr\eta_{ij3}k_1}=\MMMM^{-1}
\pmatrix{\meanemf_i^{1j}\cr\meanemf_i^{2j}},
\EN
where
\EQ
\MMMM=\pmatrix{
\cos k_1 z & -\sin k_1 z\cr
\sin k_1 z & ~~\cos k_1 z}
\EN
is the same matrix for each value of $q$ and each of the two components
$i=1,2$ of $\meanemf_i^{pq}$.
Note that $\det\MMMM=1$, so the inversion procedure is well behaved
and trivial.

Given the form of the test fields, we can 
compute the mean electromotive force 
$\meanEMF^{pq}=\overline{\uu\times{\bb}^{pq}}$ for a given
test field $\meanBB^{pq}$. 
The mean emf $\meanEMF^{pq}$ 
is computed by solving the equations
\EQ
\label{btest}
{\partial\bb^{pq}\over\partial t}=\nab\times\left(
\meanUU\times\bb^{pq} + \uu\times\meanBB^{pq}\right)+\GG^{pq}
+\eta\nabla^2\bb^{pq}
\EN
for each test field $\meanBB^{pq}$ along with the momentum equation for 
the fluctuating velocity field (see further below).\footnote{Note that in
the corresponding expression (27) of Brandenburg (2005) the $\meanUU$ term
is incorrect. This did not affect his results because $\meanUU=0$.}
Here
\EQ
\GG^{pq}=\nab\times\left(\uu\times\bb^{pq}-\overline{\uu\times\bb^{pq}}\right)
\EN
is a nonlinear term that would normally be neglected under FOSA,
but will be retained in the numerical simulations.
For sufficiently large values of $R_{\rm m}$ the small-scale field 
$\bb^{pq}$ can grow exponentially due to small-scale dynamo action. 
An important question to ask is whether the predictions of FOSA work even
in the presence of such a small-scale dynamo.

We adopt an isothermal equation of state with constant speed of sound,
$c_{\rm s}$, so the momentum and continuity equations are
\EQ
{\partial\uu\over\partial t}=-\uu\cdot\nab\uu-c_{\rm s}^2\nab\ln\rho
+\ff+\rho^{-1}\nab\cdot2\rho\nu\SSSS,
\EN
\EQ
{\partial\ln\rho\over\partial t}=-\uu\cdot\nab\ln\rho-\nab\cdot\uu,
\EN
where $\ff$ is a random forcing function consisting of circularly
polarized plane waves with positive helicity and random direction,
$\SSSS$ is the traceless rate of strain tensor.
The length of the wavevector of the forcing function,
$|\kk_{\rm f}|$, is chosen to be between 4.5 and 5.5, so the average
is around $k_{\rm f}=|\kk_{\rm f}|\approx5$.
The ratio $k_{\rm f}/k_1$ is referred to as the scale separation ratio.
It must be large enough to ensure that higher derivatives in \Eq{meanemf1}
can be ignored and that a large-scale field could grow,
if it was allowed to do so.
For fully helical turbulence, a ratio of 5 is already sufficient,
but 2.3 is not; see Fig.~23 of \cite{Hau04}.
We adjust the strength of the forcing such that the flow remains
clearly subsonic (mean Mach number is below 0.2), so for all
practical purposes the flow can be considered nearly incompressible.
The details of the forcing function used in the present work
can be found in Appendix A of \citep{BS05b}.

We do not include the Lorentz force in the momentum equation
since we want to study the mean emf in the purely kinematic limit.
We ignore here the possibility of a mean flow; such flows have not been
seen to emerge under the simple conditions considered here.
In the following we use the \textsc{Pencil Code}\footnote{
\url{http://www.nordita.org/software/pencil-code}}, where
the test field algorithm has already been implemented.
We employ periodic boundary conditions and use a resolution of up to
$512^3$ meshpoints for the run with the largest fluid Reynolds number.
We use a domain of size $(2\pi)^3$, so the smallest wavenumber is $k_1=1$.

\section{Results}

\begin{figure}\begin{center}
\includegraphics[width=\columnwidth]{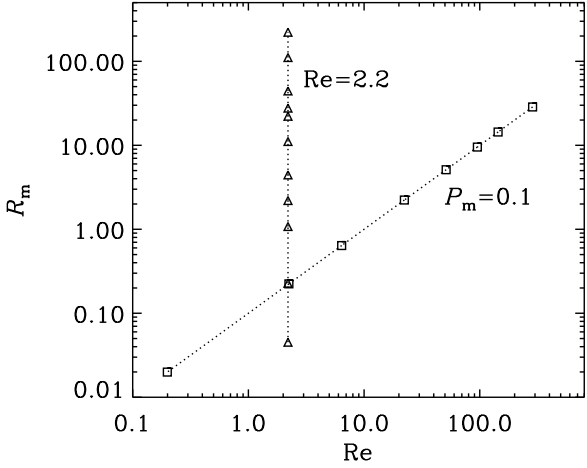}
\end{center}\caption[]{
Parameters space covered by the simulations presented in this paper.
}\label{ReRm}\end{figure}

We are particularly interested in the dependence of $\alpha_{ij}$ and
$\eta_{ij3}$ on $R_{\rm m}$ and have considered cases where
either the fluid Reynolds number was fixed,
$\mbox{Re}=u_{\rm rms}/(\nu k_{\rm f})=2.2$,
or where $\mbox{Re}=10R_{\rm m}$, corresponding to a magnetic
Prandtl number of $P_{\rm m}=\nu/\eta=0.1$; see \Fig{ReRm}.

The flow is isotropic and, not surprisingly, we find that,
to a good approximation,
the $\alpha$ and $\eta_{\rm t}$ tensors are isotropic with
\EQ
\alpha_{11}=\alpha_{22}\equiv\alpha,\quad
\eta_{123}=-\eta_{213}\equiv\eta_{\rm t},
\EN
and $\alpha_{12}=\alpha_{21}=\eta_{113}=\eta_{223}=0$.
The quantity $\eta_{\rm t}$ is simply referred to as turbulent magnetic
diffusivity.
Although the code is capable of solving for the full
$\alpha_{ij}$ and $\eta_{ij3}$ tensors, we simplify matters by solving
only for $\alpha_{i1}$ and $\eta_{i13}$ using just two test fields.

\begin{figure}\begin{center}
\includegraphics[width=\columnwidth]{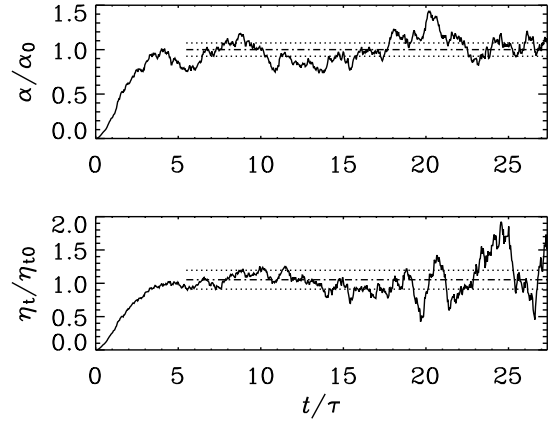}
\end{center}\caption[]{
Time series of the $z$ averaged values $\alpha$ and $\eta_{\rm t}$
for $R_{\rm m}=220$ and $\mbox{Re}=2.2$.
Time is expressed in turnover times, $\tau=(u_{\rm rms}k_{\rm f})^{-1}$.
Dash-dotted lines give the time average and dotted lines mark
error margins obtained by averaging over subsections of the
full time series (see text).
In this run there is small-scale dynamo action leading to strong
fluctuations for $t/\tau>25$, making the determinations of
reliable averages harder at late times.
}\label{alp}\end{figure}

We present the results for $\alpha$ and $\eta_{\rm t}$ normalized
to the respective expressions obtained using FOSA
for large magnetic Reynolds numbers \citep{Mof78,KR80},
\EQ
\alpha_0=-\onethird\tau\overline{\oo\cdot\uu},\quad
\eta_{\rm t0}=\onethird\tau\overline{\uu^2},
\EN
where $\tau$ denotes the correlation time of the turbulence.
Using the definitions of the Strouhal number,
\EQ
\mbox{St}=\tau u_{\rm rms}k_{\rm f}
\EN
and the fact that $\mbox{St}\approx1$ for large enough magnetic
Reynolds numbers \citep{BS05b,BS07}, we expect
\EQ
\alpha_0=-\onethird u_{\rm rms},\quad
\eta_{\rm t0}=\onethird u_{\rm rms}k_{\rm f}^{-1}
\EN
for a flow that is maximally helical and has positive helicity.
For $\Rm<1$ the relevant value of $\tau$ is no
longer the dynamical time scale, $(u_{\rm rms}k_{\rm f})^{-1}$, but the
resistive one, $(\eta_{\rm t}k_{\rm f}^2)^{-1}$.
Therefore, both $\alpha/\alpha_0$ and $\eta_{\rm t}/\eta_{\rm t0}$
have to be scaled by $R_{\rm m}$ for $R_{\rm m}<1$.

The test field procedure yields $\alpha$ and $\eta_{\rm t}$ as functions
of $z$ and $t$.
Since the turbulence is homogeneous, we average these data first over $z$
and calculate then time averages over the full time series.
An example of such a time series is shown in Fig.~\ref{alp} for $\Rey=2.2$
and $\Rm=220$.
Note that even for $\Rm\gg1$ the time averages of $\alpha/\alpha_0$ and
$\eta_{\rm t}/\eta_{\rm t0}$ are close to unity, i.e.\ the
predictions from FOSA appear to be reasonably accurate.
We use such time series to calculate error bars 
as the maximum departure between these averages
and the averages obtained from one of three equally long subsections
of the full time series.

\begin{figure}\begin{center}
\includegraphics[width=\columnwidth]{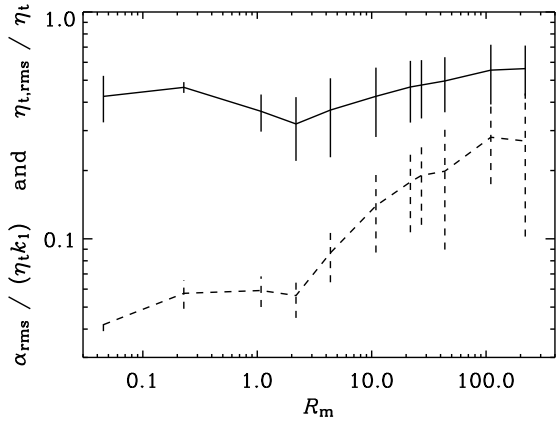}
\end{center}\caption[]{
Dependences of $\alpha_{\rm rms}$ (normalized by $\eta_{\rm t0}k_1$,
solid line), and $\eta_{\rm t,rms}$ (normalized by $\eta_{\rm t0}$,
dashed line), on the magnetic Reynolds number for $\mbox{Re}=2.2$.
The vertical bars denote the error estimated by averaging over
subsections of the full time series.
}\label{Re2_alprms}\end{figure}

The degree of fluctuations in the time series of $\alpha$ and
$\eta_{\rm t}$ is quite moderate and not at all as strong as in the
nonlinear regime where fluctuations of $\alpha$ can even dominate over
the mean value \citep{CH96}.
The latter is likely to be a consequence of a very small
(catastrophically quenched) mean value in the nonlinear regime.
On the other hand, in the kinematic regime the level of fluctuations is found
not to vary significantly with $\Rm$; see \Fig{Re2_alprms}.
A weak $\Rm$ dependence has also been found in the presence of shear
\citep{BRRK07}, where such fluctuations can contribute to dynamo
action by an incoherent $\alpha$ effect \citep{VB97}.
However, at late times, i.e.\ toward the end of the simulation,
the degree of fluctuations increases (Fig.~\ref{alp}).
This has to do with the emergence of small-scale dynamo action that
leads to the production of strong small-scale magnetic fields, $\bb^{pq}$,
although this does not affect the resulting time averaged emf for
a reasonably long stretch of time.

\begin{figure}\begin{center}
\includegraphics[width=\columnwidth]{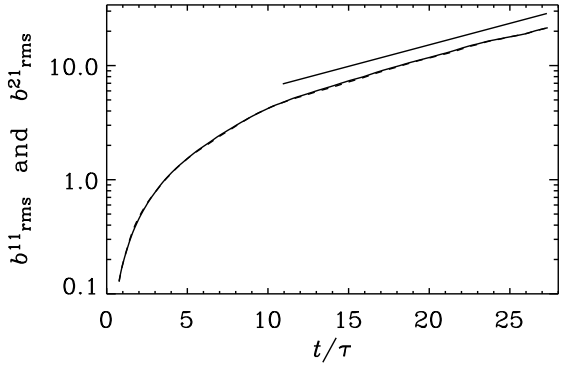}
\end{center}\caption[]{
Evolution of the root-mean-squared value of the
small-scale magnetic fields $\bb^{11}$ for $\Rey=2.2$ and $\Rm=220$.
(The result for $\bb^{21}$ is overplotted as a dashed line,
but it is almost indistinguishable from the solid line for $\bb^{11}$.)
Note the nearly exponential growth for $t/\tau>10$,
as illustrated by the straight line with a slope corresponding to
a growth rate of $1.55\tau^{-1}$.
}\label{brms}\end{figure}

In our simulations with $\Rey=2.2$, small-scale dynamo action with
exponential growth of the rms value of $\bb^{pq}$ is found when
$\Rm$ is larger than a certain critical value $R_{\rm m,cr}$ that
seems to be somewhere between 28 and 44; see Fig.~\ref{brms}
which shows the growth of the rms value of
$\bb^{pq}$ for $\Rm=220$ and $\Pm=\Rm/\Rey=100$.

\begin{figure}\begin{center}
\includegraphics[width=\columnwidth]{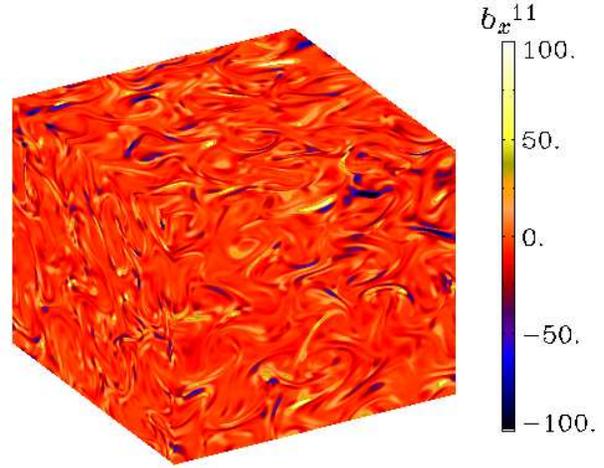}
\end{center}\caption[]{
Colour-coded (or grey-scale) representation of $b_x^{11}$ on the periphery
of the box for a run with $\Rey=2.2$ and $\Rm=220$ at $t/\tau=28$.
The colour/grey scale has been clipped at $\pm100$, even though the
extrema are at $\pm200$.
Note the extreme intermittency as evidenced by the presence of
extended nearly field-free regions.
}\label{bx11}\end{figure}

On the other hand, for $\Rm<R_{\rm m,cr}$ the 
rms value of the small-scale field settles to a constant value.
In the supercritical case the magnetic field is highly intermittent
in the sense that only in a few places the magnetic field reaches
large positive and negative values; see Fig.~\ref{bx11}.
Such intermittency is typical of large-$\Pm$ small-scale dynamo action
in the kinematic stage \citep{ZRS90,BS05a}.

\begin{figure}\begin{center}
\includegraphics[width=\columnwidth]{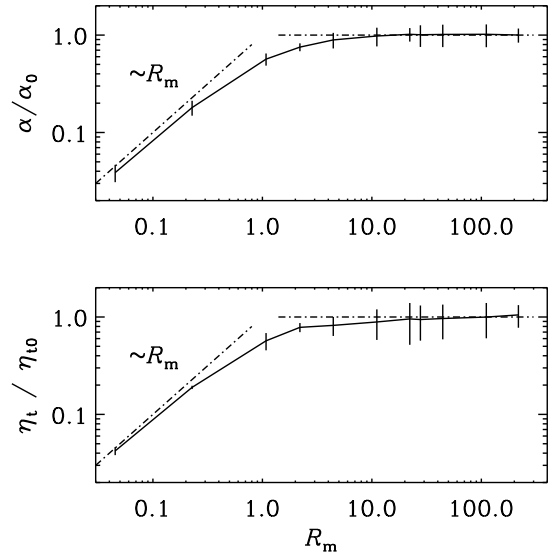}
\end{center}\caption[]{
Dependence of the normalized values of $\alpha$ and $\eta_{\rm t}$
on $R_{\rm m}$ for $\mbox{Re}=2.2$.
The vertical bars denote twice the error estimated by averaging over
subsections of the full time series (see text).
The run with $\Rm=220$ ($\Rey=2.2$) was done at a resolution of
$512^3$ meshpoints.
}\label{Re2}\end{figure}

\begin{figure}\begin{center}
\includegraphics[width=\columnwidth]{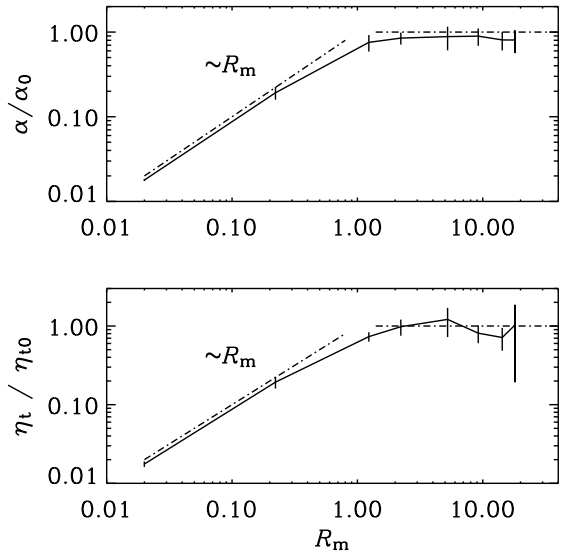}
\end{center}\caption[]{
Same as Fig.~\ref{Re2}, but for $P_m=0.1$.
The run with $\Rm\approx22$ ($\Rey\approx220$) required a resolution of
$512^3$ meshpoints.
}\label{Pm_01}\end{figure}

The corresponding normalized values of $\alpha$ and $\eta_{\rm t}$ as a
function of $\Rm$ are shown in Fig.~\ref{Re2}, for the case when $\Rey=2.2$.
\Fig{Pm_01} shows the results for simulations with $\Pm=0.1$. 
It turns out that in all cases with $\Rm>1$,
$\alpha/\alpha_0\approx\eta_{\rm t}/\eta_{\rm t0}\approx1$,
while for $\Rm<1$ these ratios are equal to $\Rm$.

Also in the case with $\Pm=0.1$, where we explore larger values of
$\Rey$ up to 300, $\alpha/\alpha_0$ and $\eta_{\rm t}/\eta_{\rm t0}$
reach values close to unity.

We may conclude that for isotropic homogeneous turbulence
the high conductivity results obtained under FOSA
are reasonably accurate up to the moderate values of $\Rm$ that we have tested.
Interestingly, this conclusion is obtained even in the presence of a
small-scale dynamo, where $\bb$ is growing exponentially.
Conversely, when different values of $\alpha$ and $\eta_{\rm t}$
are found under specific circumstances, then this must be related to
the nature of these circumstances and does not indicate the break down
of FOSA in general.

How is it then possible that the high conductivity limit of FOSA works
even though the correlation time of the velocity field is comparable
to the eddy turnover time?
A possible reason for this could be that in the kinematic regime the
high conductivity limit of FOSA gives similar predictions than the
minimal $\tau$ approximation (MTA) where the triple correlations are
not neglected, but replaced by the quadratic correlations divided by a
turnover time.
This closure assumption is not well justified, although numerical
simulations (for $\Rm\leq300$)
support some aspects of this closure \citep{BS05b,BS07}.
Let us also emphasize that in this work we have not probed two particular
aspects where FOSA and MTA depart from each other: feedback from helical
mean fields \citep[see][for a review]{BS05a}, and strong time dependence.
In the latter case a time derivative of $\meanEMF$ would be important in
giving the mean field equation the character of a wave equation
\citep{BF02,BKM04}.

\section{Conclusions}

The present work has shown that for isotropic turbulence the first
order smoothing results give quite accurate expressions for the $\alpha$
effect and the turbulent magnetic diffusivity in the kinematic regime.
This result comes almost as a surprise given that in recent years mean
field theory has been seriously challenged based on numerical simulations.
However, we can now clearly say that for
moderate Reynolds numbers up to about
220 and under the conditions stated (scale separation, isotropy, etc.)
there is {\it no} evidence that the kinematic results obtained using
FOSA are flawed, even though its applicability can then no longer be
guaranteed and we know of its shortcomings in the nonlinear regime
\citep{BS05a}.
As explained above, a possible reason for this might be that the
predictions of FOSA and MTA are rather similar, even though FOSA loses
its justification whilst MTA hinges on a not well justified closure
hypothesis.

The emergence of small-scale dynamo action for supercritical values
of $\Rm$ of about 30
does not seem to affect the average values of $\alpha$ and
$\eta_{\rm t}$.
This suggests that the exponentially growing part of the
small-scale field 
does not make a contribution to the mean emf $\meanemf$,
correlated with the imposed test fields.
However, small-scale dynamo action makes the calculation of reliable
average values of $\alpha$ and $\eta_{\rm t}$ more difficult.
Of course, in the supercritical case the long time limit of any
kinematic problem becomes unphysical.
Nevertheless, within a certain time interval the averaged values of
$\alpha$ and $\eta_{\rm t}$ match reasonably with theoretical
expectations of FOSA.

\section*{Acknowledgments}
SS and KS thank Nordita for hospitality during the course of this work.
SS would like to thank the Council of Scientific and Industrial Research,
India for providing financial support.
We acknowledge the use of the HPC facility (Cetus cluster) at IUCAA.

\newcommand{\ybook}[3]{ #1, {#2} (#3)}
\newcommand{\yjfm}[3]{ #1, {J.\ Fluid Mech.,} {#2}, #3}
\newcommand{\yprl}[3]{ #1, {Phys.\ Rev.\ Lett.,} {#2}, #3}
\newcommand{\ypre}[3]{ #1, {Phys.\ Rev.\ E,} {#2}, #3}
\newcommand{\yapj}[3]{ #1, {ApJ,} {#2}, #3}
\newcommand{\yan}[3]{ #1, {AN,} {#2}, #3}
\newcommand{\yana}[3]{ #1, {A\&A,} {#2}, #3}
\newcommand{\ygafd}[3]{ #1, {Geophys.\ Astrophys.\ Fluid Dyn.,} {#2}, #3}
\newcommand{\ypf}[3]{ #1, {Phys.\ Fluids,} {#2}, #3}
\newcommand{\yproc}[5]{ #1, in {#3}, ed.\ #4 (#5), #2}
\newcommand{\yjour}[4]{ #1, {#2} {#3}, #4.}
\newcommand{\sapj}[1]{ #1, {ApJ,} (submitted)}

\label{lastpage}
\end{document}